# Spinning twisted ribbons: when two holes meet on a curved liquid film

**Jack H.Y. Lo**[1], **Yuan Liu**[1], **Tariq Alghamdi**[2], **Muhammad F. Afzaal**[1] and **S.T. Thoroddsen**[1]

[1]Division of Physical Sciences and Engineering, King Abdullah University of Science and Technology (KAUST), Thuwal 23955-6900, Saudi Arabia

[2]Mechanical Engineering Department, College of Engineering and Architecture, Umm Al-Qura University, Makkah 21955, Saudi Arabia

**Corresponding authors:** Jack H.Y. Lo, hauyung.lo@kaust.edu.sa;
S.T. Thoroddsen, sigurdur.thoroddsen@kaust.edu.sa

The rupture of a liquid film, where a thin liquid layer between two other fluids breaks and forms holes, commonly occurs in both natural phenomena and industrial applications. The post-rupture dynamics, from initial hole formation to the complete collapse of the film, are crucial because they govern droplet formation, which plays a significant role in many applications such as disease transmission, aerosol formation, spray drying nanodrugs, oil spill remediation, inkjet printing and spray coating. While single-hole rupture has been extensively studied, the dynamics of multiple-hole ruptures, especially the interactions between neighbouring holes, are less well understood. Here, this study reveals that when two holes 'meet' on a curved film, the film evolves into a spinning twisted ribbon before breaking into droplets, distinctly different from what occurs on flat films. We explain the formation and evolution of the spinning twisted ribbon, including its geometry, orbits, corrugations and ligaments, and compare the experimental observations with models. We compare and contrast this phenomena with its counterpart on planar films. While our experiments are based on the multiple-hole ruptures in corona splash, the underlying principles are likely applicable to other systems. This study sheds light on understanding and controlling droplet formation in multiple-hole rupture, improving public health, climate science and various industrial applications.

## 1. Introduction

The rupture of freely suspended liquid films has been extensively researched, with various studies focusing on different aspects of the process. While many studies investigate the mechanisms preceding the rupture (Thoroddsen, Etoh & Takehara 2006; Vernay, Ramos & Ligoure 2015; Lo, Liu & Xu 2017; Duchemin & Josserand 2020 Oratis *et al.* 2020; Poulain & Carlson 2022; Sprittles *et al.* 2023), others focus on the mechanisms following it – notably, the rupture of a liquid film produces a large number of droplets, which can be detrimental or beneficial depending on their application (Debrégeas *et al.* 1995; Villermaux 2007; Lhuissier & Villermaux 2009; Bird *et al.* 2010; Feng *et al.* 2014; Villermaux 2020; Jiang *et al.* 2022, 2024). This is crucial in various fields, including disease transmission, aerosol formation, spray drying nanodrugs, oil dispersal, inkjet printing and spray coating, which in turn affect health, climate and many industrial applications.

Film rupture can be classified into two categories: single-hole rupture and multiple-hole rupture. In a single-hole rupture, only one hole forms within the time scale of the film's complete collapse (Debrégeas *et al.* 1995; Bird *et al.* 2010; Feng *et al.* 2014; Jiang *et al.* 2022). This scenario is often observed when an external object punctures the film, such as bursting a bubble with a needle. In contrast, multiple-hole rupture involves the formation of several holes either simultaneously or in close succession. This typically occurs in thin, unstable films, where holes nucleate spontaneously at different points. Multiple-hole rupture is likely when the liquid contains a dispersed phase, such as emulsions, air bubbles or solid suspensions. It is also likely to occur when the liquid film is expanding and thus thinning rapidly. Examples include bag breakup of a falling raindrop (Villermaux & Bossa 2009), rapid expansion of a bubble (Vledouts *et al.* 2016), bursting of a surface bubble (Qian *et al.* 2023), corona splash (Thoroddsen *et al.* 2006; Aljedaani *et al.* 2018), fan spray nozzle (Dombrowski & Fraser 1954; Lhuissier & Villermaux 2013), drop impact on a small surface (Vernay *et al.* 2015), drop impact on a superhydrophobic substrate (Kim *et al.* 2020), and the trapped air film of drop impact onto a pool (Thoroddsen *et al.* 2012) or solid surfaces (Li, Vakarelski & Thoroddsen 2015; Langley *et al.* 2018).

The multiple-hole rupture cannot be explained simply as a superposition of single-hole ruptures, because the holes interact with one another. Building on earlier studies (Dombrowski & Fraser 1954; Lhuissier & Villermaux 2013), significant progress has been made recently in understanding hole–hole interactions on planar films (Néel *et al.* 2020; Agbaglah 2021; Tang, Adcock & Mostert 2024). When two holes collide at high Weber number, a transverse lamellar sheet emerges and breaks into smaller droplets in a process known as rim splashing (Néel *et al.* 2020; Tang *et al.* 2024). The droplet formation is closely related to the ligament growth, which has been studied in detail recently on expanding liquid sheets (Wang *et al.* 2018; Wang & Bourouiba 2021).

While the mechanism that triggers the film rupture is not the focus of this study, it is an interesting topic that remains under active research and is likely to vary depending on specific circumstances. Possible explanations include, but are not limited to, the Rayleigh–Taylor instability, turbulence within the film, presence of tiny air bubbles trapped inside, and collisions with surrounding fine droplets of lower surface tension (Thoroddsen *et al.* 2006; Vledouts *et al.* 2016; Aljedaani *et al.* 2018; Bang *et al.* 2023; Stumpf *et al.* 2023). The probability of hole formation in turbulence-triggered rupture has been previously derived (Bang *et al.* 2023; Stumpf *et al.* 2023).

In this study, we reveal a phenomenon that is unique to curved films: when two holes meet, the liquid film evolves into a spinning twisted ribbon, as shown in figure 1(*a*–*c*) and supplementary movies 1 and 2 available at https://doi.org/10.1017/jfm.2025.10299, and then droplets are ejected due to the spinning. The ribbon is a tiny helicoid-like structure ($\sim$100 μm wide) that spins at a high speed ($\sim$5000 Hz). It can, therefore, be easily overlooked without careful inspection. Nevertheless, this phenomenon may have been observed as early as 1954 by Dombrowski & Fraser (1954), who wrote that 'At the instant before coalescence of the two rims, the ribbon of liquid between them may twist'. To the best of our knowledge, however, the twisted ribbon has not been studied until now. Besides the meeting of two holes, the twisted ribbon is a general feature that also emerges in various similar systems, such as the meeting of two edges (figure 1*d* and supplementary movie 3) and the spikes in corona splash (figure 1*e* and supplementary movie 4). The analysis is based on experiments of multiple-hole rupture in corona splash; we explain the formation and evolution of spinning twisted ribbons. The underlying principles are likely applicable to other systems where twisted ribbons appear.

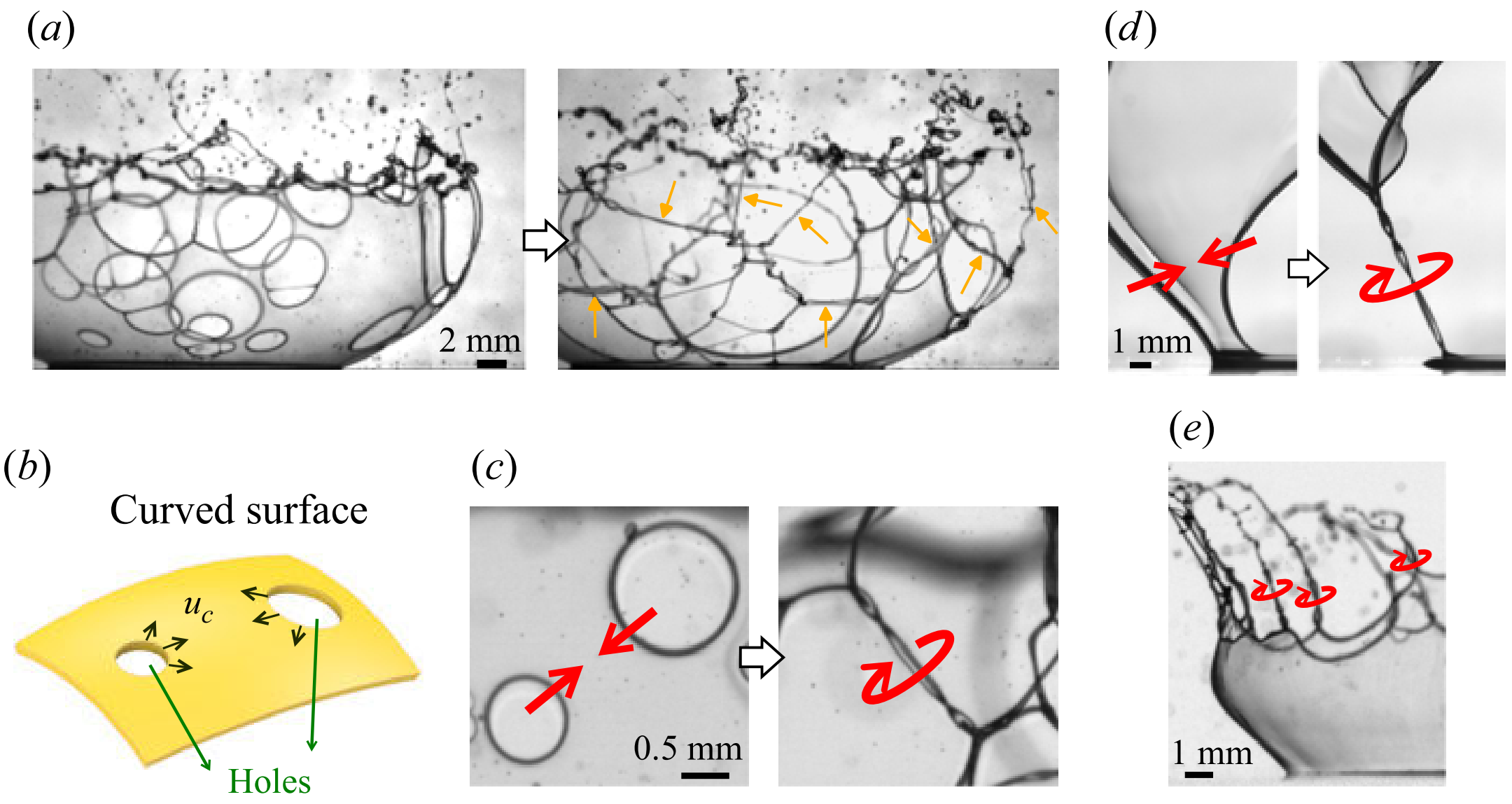

Figure 1. Examples of spinning twisted ribbons appearing on rupturing curved liquid sheets under various scenarios: (*a*–*c*) meeting of two expanding holes, (*d*) meeting of two edges, (*e*) spikes of the crown splash. See supplementary movies 1–4. (*b*,*c*) A drawing and magnified view for the case of (*a*) showing two holes expanding at a constant speed $u_c$. In (*d*) and supplementary movie 3, the highest observed spinning frequency is 5200 Hz.

## 2. Experimental methods

We observe the ruptures appearing in the corona splash, induced by the high-speed impact of viscous drops on glass slides coated with a thin lower-viscosity liquid film. In corona splash, the spreading liquid sheet lifts upwards away from the substrate to form a 'crown', which is essentially a curved thin film, as shown in figure 1(*a*) and supplementary movie 1. We focus only on the film ruptures, rather than on the splashing as a whole, which is a complex phenomenon still under active research (Thoroddsen *et al.* 2006; Yarin 2006; Aljedaani *et al.* 2018; Sanjay *et al.* 2023; Sykes *et al.* 2023; Khan, Jin & Yang 2024; Tian *et al.* 2024).

The drop shape is flattened due to aerodynamics stress at high falling speeds. The drops impact at speeds $U = 6.4$–$8.1\ \mathrm{m\,s^{-1}}$, with vertical diameters $D_V = 2.3$–$3.7$ mm and horizontal diameters $D_H = 3.5$–$4.8$ mm. The drops are prepared using either silicon oil or glycerol–water mixtures, with viscosities of 30–50 cSt and surface tensions of 21 or 65 $\mathrm{mN\,m^{-1}}$. The corresponding ranges of the drop-impact Weber, Reynolds and Ohnesorge numbers are $We = \rho D U^2/\gamma = 5170-12\,000$, $Re = \rho D U/\mu = 400-1200$ and $Oh = \mu/\sqrt{\rho D \gamma} = 0.09{-}0.19$, respectively, where $\rho$, $\gamma$, $\mu$ and $D$ are the density, surface tension, viscosity and volume-equivalent diameter of the drop.

The coated liquid film on the glass slide consists of either 1.5 cSt silicon oil with a thickness of 53 $\mu$m or ethanol with a thickness of 35 $\mu$m. The thickness is calculated from the volume of the deposited liquid and the area of the substrate. As the film coated on the solid substrate is very thin, following the original spray, the crown of the corona splash (i.e. the curved film) originates from the liquid in the drop, rather than from the liquid film coated on the glass slide (Thoroddsen *et al.* 2006). This is verified in our experiments by dying the drop, as presented in Appendix A. Therefore, the liquid properties of the substrate-coated film do not affect the ruptures directly.

The crown of the corona splash is a curved liquid thin film. The film thickness $\delta$ is deduced from the measured speed by inverting the Taylor–Culick relation. The film thickness varies slightly across different samples as the crown expands and thins over time, with a mean of $7.8 \pm 1.3$ $\mu$m for the silicone oil and a range of 10–40 $\mu$m for the glycerol–water mixture.

We record the impact processes using one or two high-speed cameras, with a frame rate of 31 000 frames per second and a resolution of 14.5 $\mu$m pixel$^{-1}$. On a curved surface, the plane of ruptures may not be parallel to the image plane of the camera, leading to parallax errors in length (Stumpf *et al.* 2023). We correct the parallax error on the curved surface by using the method outlined in Appendix B. To focus on the dynamics of the ruptures, we present the data in a moving inertial frame that offsets the linear motion of the liquid film.

## 3. Spinning twisted ribbons

### 3.1. *Phenomenology*

To describe how spinning twisted ribbons are formed, we begin by considering a curved liquid sheet that has developed two holes, as illustrated in figure 2(*a*). It is well known that the holes expand at a constant speed, known as the Taylor–Culick velocity, given by $u_c = \sqrt{2\gamma/(\rho\delta)}$, where $\gamma$ is the surface tension, $\rho$ is the density of the liquid, and $\delta$ is the thickness of the sheet. Because the radius of curvature of the crown film, $R_f$, is much larger than the travelled distance of the rim, $d$, it has negligible impact on the measured Taylor–Culick velocity $u_c$. Specifically, in our experiments, $d/R_f < 0.1$, so that the projection factor $\cos\theta_p \approx 1 - ((d/R_f)^2/2) \approx 1$, where $\theta_p$ is the angle between the image plane and the tangential plane at the hole. As a hole expands, the displaced fluid accumulates at the circular rim of the liquid sheet, while the thickness of the rest of the sheet remains constant (Savva & Bush 2009).

As the two holes expand, their edges will eventually 'meet', while the subsequent development depends on whether the liquid sheet is planar or curved. On a planar liquid sheet, it has been shown that the rims will collide head-on, producing lamella and ejecta when the collision Weber number is sufficiently high (Néel *et al.* 2020; Tang *et al.* 2024). On a curved liquid sheet, in contrast, the rims cross each other laterally, as demonstrated in supplementary movie 2 and figure 2(*a*,*b*). This occurs because the trajectories of the rims deviate from the initial curved surface: since surface tension acts tangentially, initially, there is no centripetal force to maintain the rims on a curved path, as illustrated in figure 2(*a*) and supplementary movie 5. Note that at later times, as the rims deviate from the initial surface, the liquid film bends outwards, and surface tension provides centripetal acceleration for curvilinear motions. In the case of a single-hole rupture on a bubble, it has been observed that the centripetal acceleration destabilises the rim via Rayleigh–Taylor instability (Lhuissier & Villermaux 2012; Jiang *et al.* 2022). However, in the current case of multiple-hole ruptures, the rims meet one another (∼0.3 ms) before the instability becomes prominent (∼50 ms) (figure 19 of Lhuissier & Villermaux 2012).

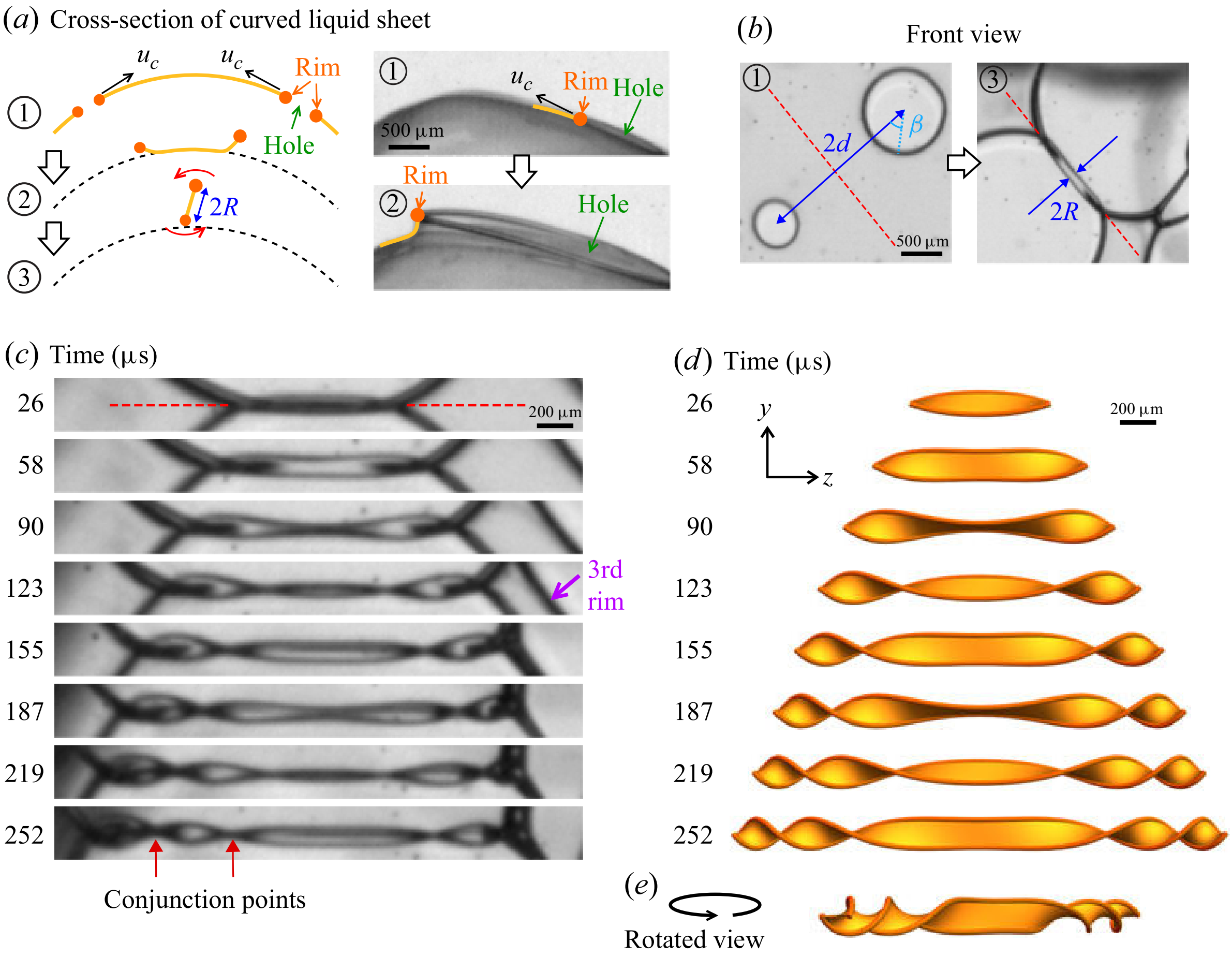

Figure 2. The formation mechanism of the spinning twisted ribbon. (*a*) Considering a curved liquid sheet with two holes that have punctured at slightly different times. Their rims expand at the Taylor-Culick velocity $u_c$, as indicated by the arrows. The trajectories of the rims deviate from the initial curved surface because the centripetal force is insufficient to keep the rims on track. The circled numbers indicate the sequence of events. A single-hole example is given here and in supplementary movie 5. (*b*) Consequently, the rims cross each other laterally and begin spinning. See also supplementary movie 2. The distance between the two rupture centres is $2d$. The rotation radius is $R$. The azimuthal angle $\beta$ is measured from the centre of the hole. The rotation axis is denoted by the red dashed line. (*c, d*) Magnified video frames of the spinning twisted ribbon and the corresponding calculated surfaces, showing the region of interest along the rotation axis (red dashed line in (*b*)). The surfaces are plotted by the model (3.2), using the measured parameters $u_c = 2.69$ m s$^{-1}$, $d = 1.08$ mm, $\omega_0 = 32\,500$ s$^{-1}$ or 5170 Hz. No fitting parameters are involved. See also supplementary movie 6. (*e*) Rotated view of the last plotted ribbon surface in (*d*).

The spinning twisted ribbon is formed after the rims cross laterally. Due to the attractive surface tension of the connecting liquid sheet, the rims rotate around the axis where they cross laterally, indicated by the dashed red line in figure 2(*b*,*c*), forming the spinning twisted ribbon, as shown in supplementary movie 2 and figure 2(*c*). We define $t = 0$ as the moment that the rims cross laterally for the first time.

Note that not every pair of neighbouring holes produces a twisted ribbon. We find that the average number of ribbons per hole is $0.8 \pm 0.1$, based on a total count of 198 holes across 21 drop-impact experiments. Qualitatively, even on a curved film, head-on collisions between rims can occur in the following scenarios, preventing the ribbon formation. First, when the two holes are very close to each other, the effect of curvature is insignificant. Second, the two holes rupture simultaneously, resulting in a mirror symmetric system. This symmetry leads to an angled head-on collision.

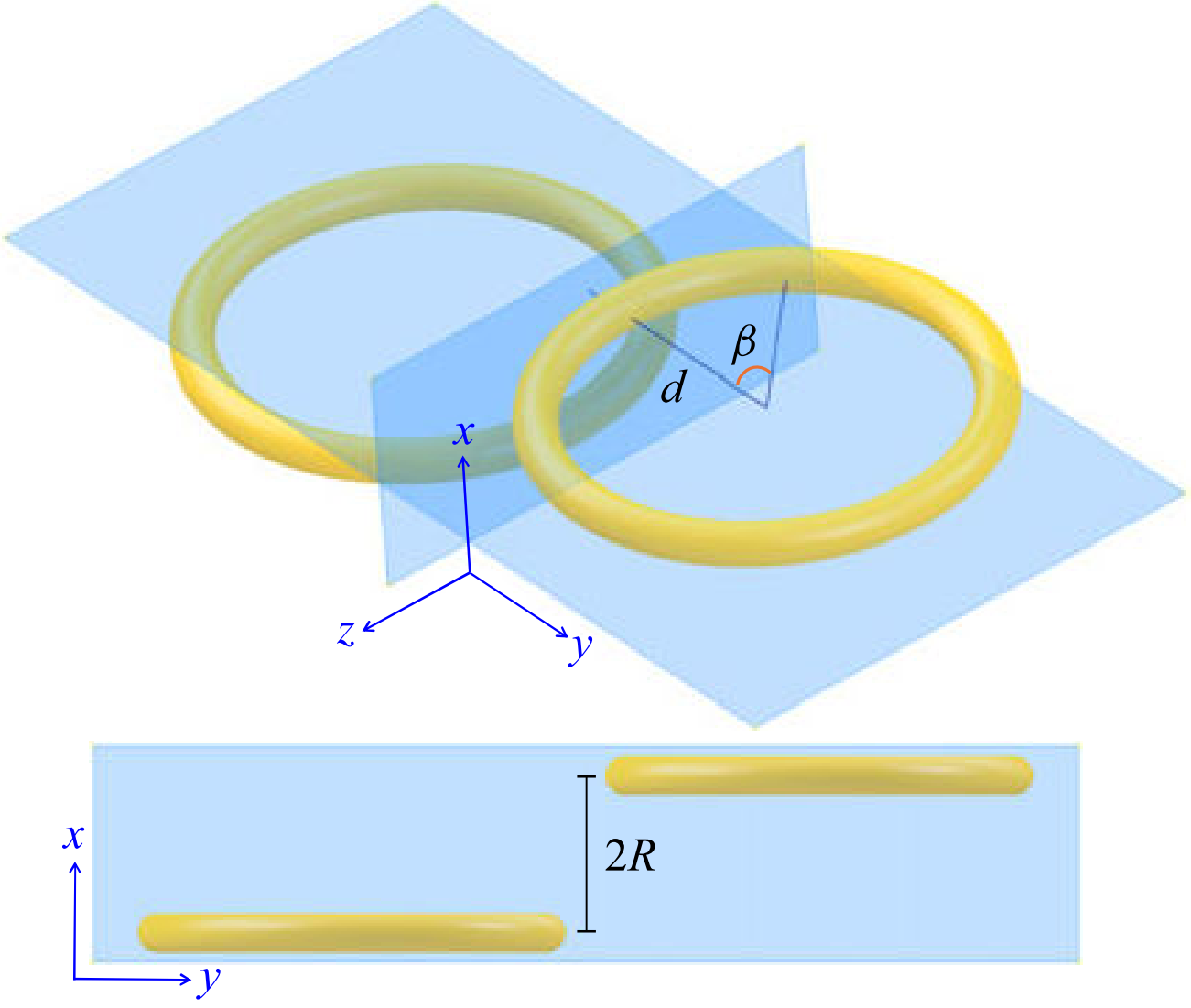

Figure 3. Schematic drawings for the derivation of (3.2).

We identify several key features of the twisted ribbon from the video images. The ribbon looks similar to a helicoid, but strictly speaking it is not. The ribbon exhibits mirror symmetry with respect to the central point. There are several points of conjunction (red arrows in figure 2*c*), where the two rims align along the same line of sight, visually overlapping but not physically touching. The number of conjunction points increases over time, indicating more twisting of the sheet, with the points moving outwards along the axis.

### 3.2. *Kinematic model*

To explain the key features of the spinning twisted ribbon, we propose a kinematic model with several idealised assumptions. The predicted structures from the model are shown in figure 2(*d*) and supplementary movie 6, and compared with the data in figure 2(*c*). The parameters involved are defined in figure 2(*a*,*b*) and elaborated by the schematic drawings in figure 3.

In this kinematic model, we consider two horizontal circular holes of equal size. Let the origin of our coordinate system be the midpoint between the rupture points (centres) of the two holes. The rims of the holes expand horizontally at speed $u_c$ from their rupture points along two planes parallel to the $yz$-plane and separated by a distance $2R$. The distance between the $xz$-plane and rupture points is $d$. At time $t = 0$, the rims reach the $xz$-plane and conjunct (relative to an observer at $x \to \infty$) for the first time. Thus we call the $xz$-plane the conjunction plane.

Next, we consider an arbitrary segment of the rim, represented by an azimuthal angle $\beta$, as shown in figure 3. It reaches the conjunction plane at time $\tau = d/(u_c \cos\beta) - d/u_c$ and at $z$-position $z_0 = d\tan\beta = (u_c\tau + d)\sin\beta$. After reaching the conjunction plane, it rotates around a rotation axis ($z$-axis) with rotation radius $R$, angular speed $\omega = u_c \cos\beta/R$ and axial speed $u_z = u_c \sin\beta$. For simplicity, here we take the rotation radius $R$ as a constant, implying a circular closed orbit. The actual non-circular open orbit is presented later with the central force model. Therefore, the motion of a segment of the rim is described by

$$
\begin{aligned}
x &= R\cos\left(\omega\left(t-\tau\right)\right),\\
y &= R\sin\left(\omega\left(t-\tau\right)\right),\\
z &= u_z(t-\tau) + z_0.
\end{aligned}
\tag{3.1}
$$

Recall that $\omega$, $\tau$, $u_z$, $z_0$ can be expressed in terms of the variable $\beta$ and constants $u_c$, $R$, $d$.

Finally, we assume that the ribbon surface is a ruled surface formed by connecting two rims by introducing a parameter $\alpha$, and each rim is a collection of rim segments described by (3.1) with parameter $\beta$ and constants $u_c$, $R$, $d$. The surface of the twisted ribbon is described by the parametric equations

$$\left.\begin{aligned} x(\alpha, \beta) &= \alpha R \cos\left(\frac{u_c t + d}{R}\cos\beta - \frac{d}{R}\right), \\ y(\alpha, \beta) &= \alpha R \sin\left(\frac{u_c t + d}{R}\cos\beta - \frac{d}{R}\right), \\ z(\beta) &= (u_c t + d)\sin\beta, \end{aligned}\right\} \tag{3.2}$$

where $-1 \leqslant \alpha \leqslant 1$, $-\beta_0 \leqslant \beta \leqslant \beta_0$. Here the range of $\beta$ is bounded by $\beta_0$ that satisfy $t \geqslant \tau$ in (3.1), given by $\cos\beta_0 = d/(u_c t + d)$.

This surface is plotted in figure 2(*d*) using measured parameters $u_c = 2.69\ \mathrm{m\,s^{-1}}$, $d = 1.08$ mm and $R = 82.8\ \mu\mathrm{m}$, obtained from the video frames in figure 2(*c*). The velocity $u_c$ and distance $d$ are measured with correction for parallax error, as described in Appendix B. The radius $R$ is obtained by measuring the central angular frequency $\omega_0 \equiv u_c/R = 32\,500\ \mathrm{s^{-1}}$ or 5170 Hz. The calculated surfaces resemble the ribbons observed in the experiment, reproducing the key features discussed earlier. No fitting parameters are involved. Note that the rims' surfaces are not included in the equations, and only added in figure 2(*d*,*e*) for the sake of clarity. The model only considers two holes, while in the experiments more than two holes frequently occur, affecting the shape of the twisted ribbon. For example, see the third rim on the right-hand side in figure 2(*c*).

Furthermore, to verify the kinematic model, we compare the positions of the measured and calculated conjunction points at different times in figure 4. By (3.1) and (3.2), the $z$-position of conjunction points are given by $z_n = (u_c t + d)\sin\beta_n$, where the angles $\beta_n$ have to satisfy the condition $\omega(t - \tau) = n\pi$ with $n = 0, 1, 2, \ldots$. Simplifying, the $z$-positions of conjunction points are

$$\frac{z_n}{d} = \sqrt{\left(\frac{u_c t}{d} + 1\right)^2 - \left(\frac{n\pi R}{d} + 1\right)^2}, \tag{3.3}$$

for $t \geqslant n\pi R/u_c$ and $n = 0, 1, 2, \ldots$. They are plotted in figure 4 using the same values of $d$ and $R$ employed in plotting figure 2(*d*). The measured data and calculated results show reasonably good agreement.

New conjunction points emerge at regular time intervals of $\pi R/u_c$. By (3.3), the speeds of conjunction points are

$$\frac{1}{u_c}\frac{\mathrm{d}z_n}{\mathrm{d}t} \approx 1 + \frac{1}{2}\left(\frac{n\pi R/d + 1}{u_c t/d + 1}\right)^2 + \cdots. \tag{3.4}$$

At large time, their speed approaches a constant value $u_c$, which is the Taylor–Culick speed in our case.

Next, we calculate the mean curvature of the ribbon by

$$H = \frac{eG - 2fF + gE}{2(EG - F^2)}, \tag{3.5}$$

where $H$ is the mean curvature, $\{E, F, G\}$ and $\{e, f, g\}$ are the coefficients of the first and second fundamental form of the surface described by (3.2). We get

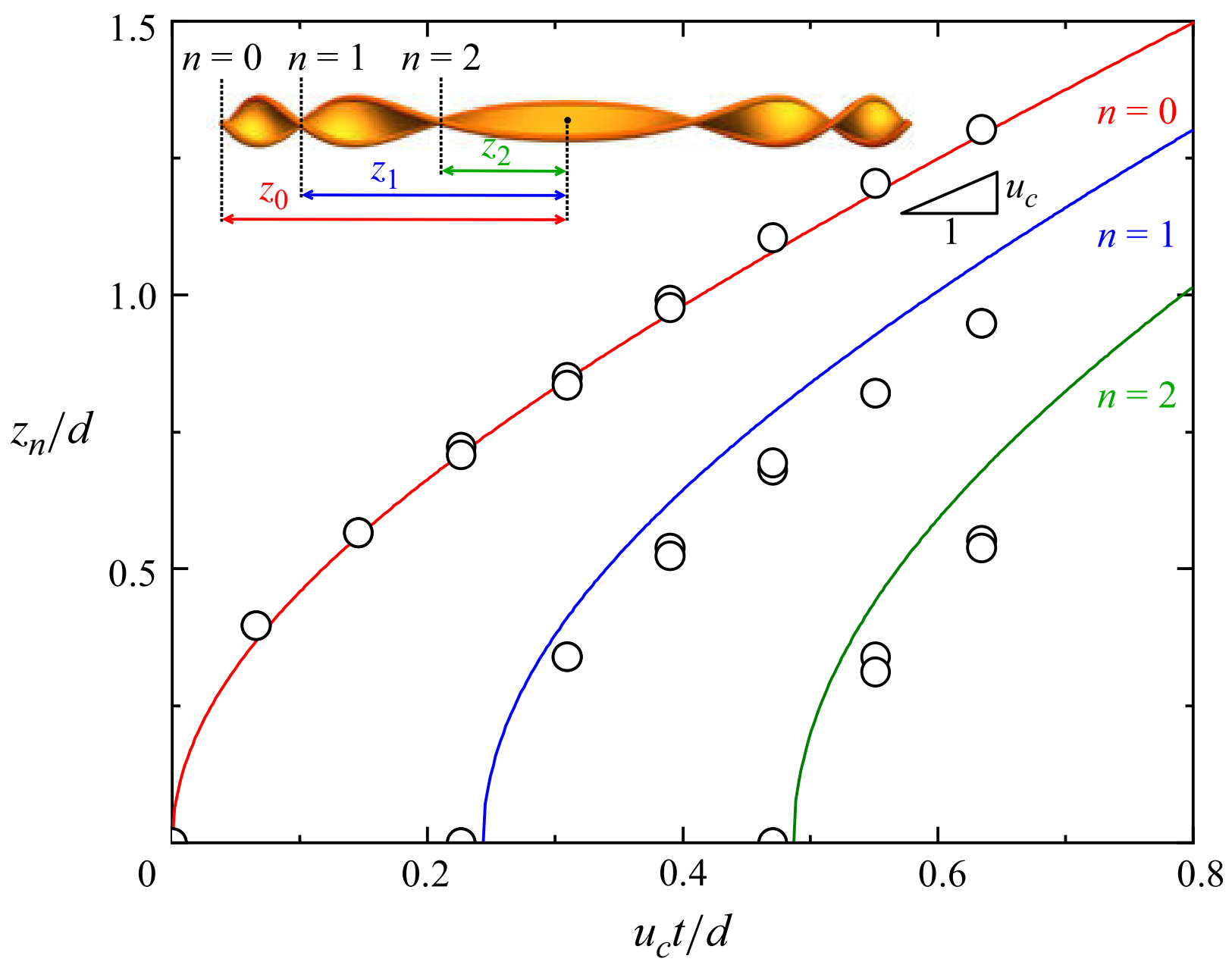

Figure 4. Positions of conjunction points at different times. The measured data (dots) agree well with the calculation results (curved lines) predicted by the kinematic model and (3.3). Some data points overlap due to the mirror symmetry with respect to the $xy$-plane.

$$H(\alpha, \beta) = \left(\frac{1}{u_ct + d}\right)\left(\frac{\alpha}{2\left(\cos^2\beta + \alpha^2 \sin^2\beta\right)^{3/2}}\right). \tag{3.6}$$

Numerically, in the range of interest $|\beta| < \pi/4$ and $|\alpha| \leqslant 1$,

$$H < \frac{0.55}{u_ct + d} < \frac{1}{d}. \tag{3.7}$$

This means the upper bound is $1/d$. Typically, the hole–axis distance $d$ (∼1 mm) is much larger than the rotation radius $R$ (∼0.1 mm), which represents the size of the ribbon. Therefore, the calculated mean curvature of the ribbon is small.

The small mean curvature obtained agrees with our expectations. First, physically, it suggests a pressure equilibrium across the film, which is known to be established rapidly after film rupture (Bird *et al.* 2010). Second, geometrically, it aligns with the visual similarity between the twisted ribbon and the helicoid, which is a minimal surface.

### 3.3. *Asymmetric kinematic model*

We extend the kinematic model to account for asymmetric cases where the sizes of the two holes are different. This scenario happens when a larger hole, which is formed earlier, interacts with a smaller hole. When the holes are the same size, it is obvious that their rims meet along a straight line at the midpoint. When the hole sizes are different, their rims meet along a hyperbola, as illustrated in figure 5. This is because the difference in radius between the two expanding holes is a constant. This is analogous to the interference of two circular waves (Pain 2005, p. 356). The hyperbola is described by the equation in polar coordinate as

$$r_h(\beta) = \frac{d(1 - e^{-2})}{e^{-1} + \cos\beta}, \tag{3.8}$$

where $e = (d_1 + d_2)/(d_2 - d_1)$ is the eccentricity, $2d = d_1 + d_2$ is the distance between the centres of the two holes, $d_{1,2}$ are the shortest distances from the centres of the holes to the ribbon and $d_1 < d_2$, as defined in figure 5(*a*). If the system is symmetric, $e^{-1} = 0$ and thus $r_h = d/\cos\beta$ becomes a straight line as expected.

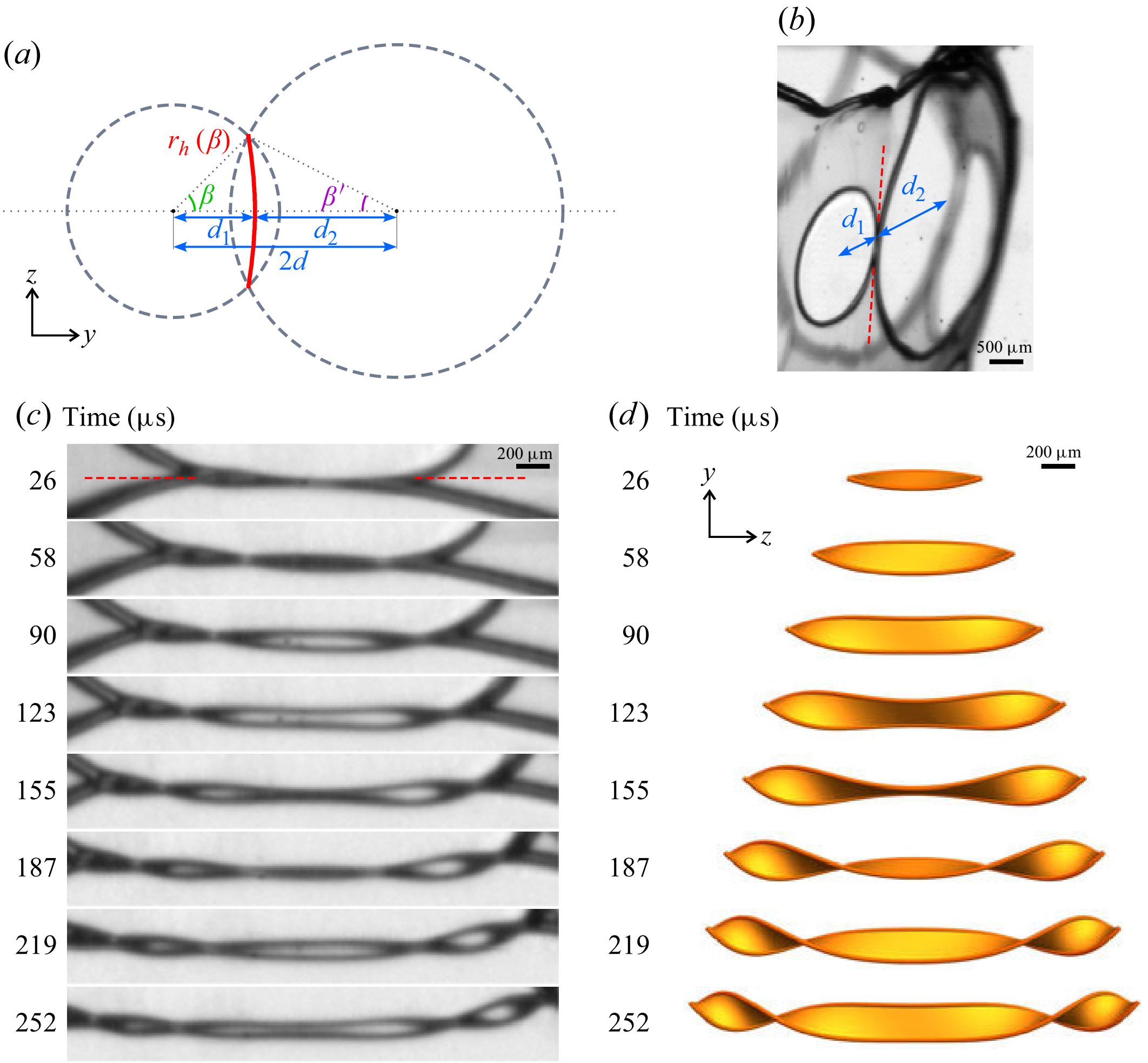

Figure 5. The asymmetric model. (*a*) Schematic diagram consisting of two holes of different sizes (dotted circles), with their rims meeting along a hyperbola (red) and forming a ribbon. (*b*) Snapshot of a small hole (left) interacting with a large hole (right). (*c*,*d*) Magnified video frames and the corresponding predicted surfaces of the spinning twisted ribbon formed by a small hole (top) and a large hole (bottom). The surfaces are plotted by (3.14), using the measured parameters $u_c$ = 2.17 m s$^{-1}$, $d_1$ = 1.19 mm, $d_2$ = 1.92 mm, $\omega_0 = 19\,500$ s$^{-1}$ or 3100 Hz.

In the asymmetric case, the momenta of the two rims are also different. The rim of the larger hole is thicker due to the larger hole size and volume conservation, as the rim collects the film liquid during its motion. Its normal component of the velocity is also larger because $\beta' < \beta$, as shown in figure 5(*a*). Therefore, the rim of the larger hole should have a larger momentum than that of the smaller hole, driving the ribbons to drift in one direction. Nevertheless, in our experiments, the liquid sheet itself is also moving, making it difficult to measure this momentum mismatch directly.

We now express the ribbon surface for the asymmetric case in a form similar to the symmetric case given in (3.2). For simplicity, we assume that their rim thicknesses are the same, so that

$$\omega = \frac{u_c}{R}\frac{\cos\beta + \cos\beta'}{2} = f_c(\beta)\omega_s,$$
$$u_z = u_c\frac{\sin\beta + \sin\beta'}{2} = f_s(\beta)u_c\sin\beta, \tag{3.9}$$

where $\omega_s = u_c \cos\beta/R$ is the angular speed for the symmetric case, $f_c(\beta)$ and $f_s(\beta)$ are geometric factors that account for the asymmetry, given by

$$f_c(\beta) = \frac{e^{-1}\cos\beta + 1}{2e^{-1}\cos\beta + r_h(\beta)\cos\beta/d}, \quad f_s(\beta) = \frac{e^{-1} + r_h(\beta)/d}{2e^{-1} + r_h(\beta)/d}. \tag{3.10}$$

If the configuration is symmetric ($e^{-1} = 0$), $f_c(\beta) = f_s(\beta) = 1$ as expected. The phase lag $\tau$ for the asymmetric case is

$$\tau = \frac{r_h(\beta)}{u_c} - \frac{r_h(0)}{u_c} = g(\beta)\tau_s, \tag{3.11}$$

where $\tau_s = d/(u_c\cos\beta) - d/u_c$ is the phase lag of the symmetric case, the geometric factor $g(\beta)$ is given by

$$g(\beta) = \frac{(1 - e^{-1})\cos\beta}{e^{-1} + \cos\beta}. \tag{3.12}$$

Therefore, similar to (3.2), the parametric equations are

$$\begin{aligned} x(\alpha, \beta) &= \alpha R \cos\left(f_c\omega_s\left(t - g\tau_s\right)\right), \\ y(\alpha, \beta) &= \alpha R \sin\left(f_c\omega_s\left(t - g\tau_s\right)\right) - y_0(\beta), \\ z(\beta) &= f_s u_c \sin\beta(t - g\tau_s)) + z_0(\beta), \end{aligned} \tag{3.13}$$

or, equivalently,

$$\begin{aligned} x(\alpha, \beta) &= \alpha R \cos\left(\left(\frac{u_c t + dg(\beta)}{R}\cos\beta - \frac{dg(\beta)}{R}\right) f_c(\beta)\right), \\ y(\alpha, \beta) &= \alpha R \sin\left(\left(\frac{u_c t + dg(\beta)}{R}\cos\beta - \frac{dg(\beta)}{R}\right) f_c(\beta)\right) - y_0(\beta), \\ z(\beta) &= f_s(\beta)\sin\beta\left(u_c t - g(\beta)\left(\frac{d}{\cos\beta} - d\right)\right) + z_0(\beta), \end{aligned} \tag{3.14}$$

where $y_0(\beta) = r_h(\beta)\cos\beta - r_h(0)$ and $z_0(\beta) = r_h(\beta)\sin\beta$. An example involving two holes of different sizes is shown in figure 5(*b*,*c*). The predicted ribbon surface is plotted in figure 5(*d*) using measured parameters $u_c = 2.17\ \mathrm{m\,s^{-1}}$, $d_1 = 1.19$ mm, $d_2 = 1.92$ mm and $R = 111$ µm. From $d_1$ and $d_2$, we get $d = 1.55$ mm and eccentricity $e = 4.24$. The calculated surfaces resemble the ribbons observed in the experiment.

In this case, $d_2/d_1 = 1.62$. For comparison, the earlier case presented in figure 2 has $d_2/d_1 = 1.07$, and its slight asymmetry can therefore be neglected.

### 3.4. *Dynamics: two-body central force model*

We elucidate the 'orbit' of the rims of the spinning ribbon by using the classical two-body central force model and direct measurements (Goldstein, Poole & Safko 2001). We assumed a circular orbit in the last section, but in fact it is open and non-circular. Consider a thin strip of the ribbon with a dumbbell-shaped cross-section, as shown in figure 6(*a*), where two circular rims are connected by a thin liquid string. The surface tension on the liquid string exerts an attractive central force between the two circular rims. For simplicity, we consider a thin strip at the centre plane ($z = 0$).

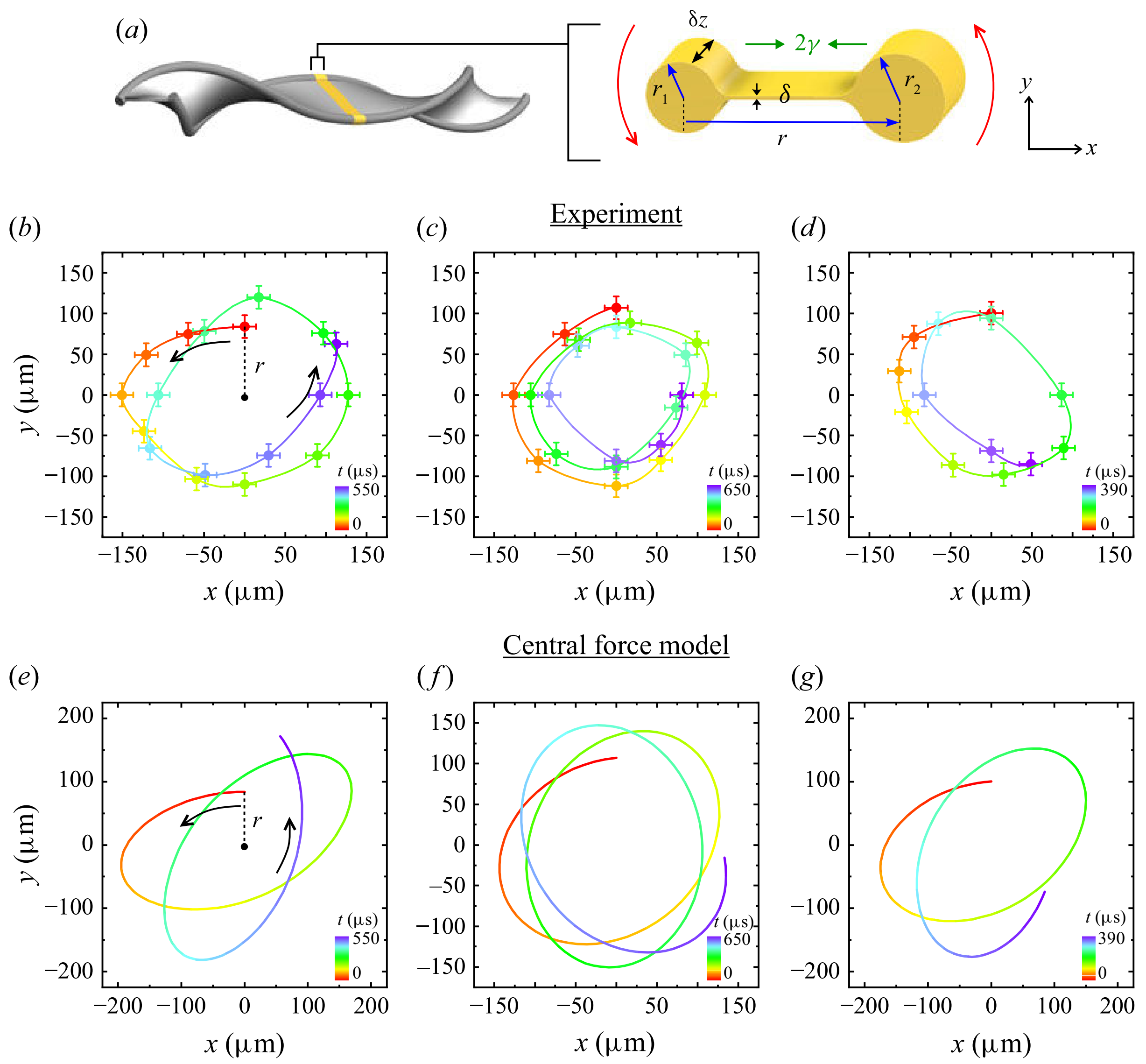

Figure 6. Explanation of the ribbon's orbit by the two-body central force model. (*a*) Sketch of a thin strip of the ribbon with a dumbbell-shaped cross-section. Two circular rims are connected by a thin liquid string, which exerts an attractive central force. (*b–d*) Measured orbits, showing the data (dots) and interpolation (line). The colours indicate time progression, ranging from red ($t=0$) to purple ($t=550$, 650 and 390 μs). (*e–g*) The corresponding orbits calculated by solving (3.15) with liquid properties $\gamma = 20.8\ \mathrm{mN\,m^{-1}}$, $\rho = 960\ \mathrm{kg\,m^{-3}}$, and initial conditions approximated from measurements: (*e*) $r(0) = 841\ \mu\mathrm{m}$, $v_\theta(0) = 4.45\ \mathrm{m\,s^{-1}}$, $v_r(0) = 0$, $m_\mu = 586\ \mu\mathrm{g\,m^{-1}}$; (*f*) $r(0) = 1071\ \mu\mathrm{m}$, $v_\theta(0) = 4.01\ \mathrm{m\,s^{-1}}$, $v_r(0) = -0.40\ \mathrm{m\,s^{-1}}$, $m_\mu = 449\ \mu\mathrm{g\,m^{-1}}$; (*g*) $r(0) = 1003\ \mu\mathrm{m}$, $v_\theta(0) = 4.94\ \mathrm{m\,s^{-1}}$, $v_r(0) = -0.33\ \mathrm{m\,s^{-1}}$, $m_\mu = 397\ \mu\mathrm{g\,m^{-1}}$. The data agree with the model semi-quantitatively.

Following the classical treatment, the two-body system is reduced to a one-body system with a relative position $\mathbf{r}(t)$ and a reduced mass per length $m_\mu = \rho\pi r_1^2 r_2^2/(r_1^2 + r_2^2)$, where $\rho$ is the density of the liquid sheet, $r_{1,2}$ are the radii of the rims. The magnitude of the central force per length is twice the surface tension $\gamma$, accounting for both the upper and lower surfaces of the liquid string. The orbit is obtained by solving the following differential equations numerically by Mathematica (Taborek 2010),

$$\begin{aligned} r''(t) - r(t)\theta'(t)^2 + \frac{2\gamma}{m_\mu} &= 0, \\ r(t)\theta''(t) + 2r'(t)\theta'(t) &= 0, \end{aligned} \tag{3.15}$$

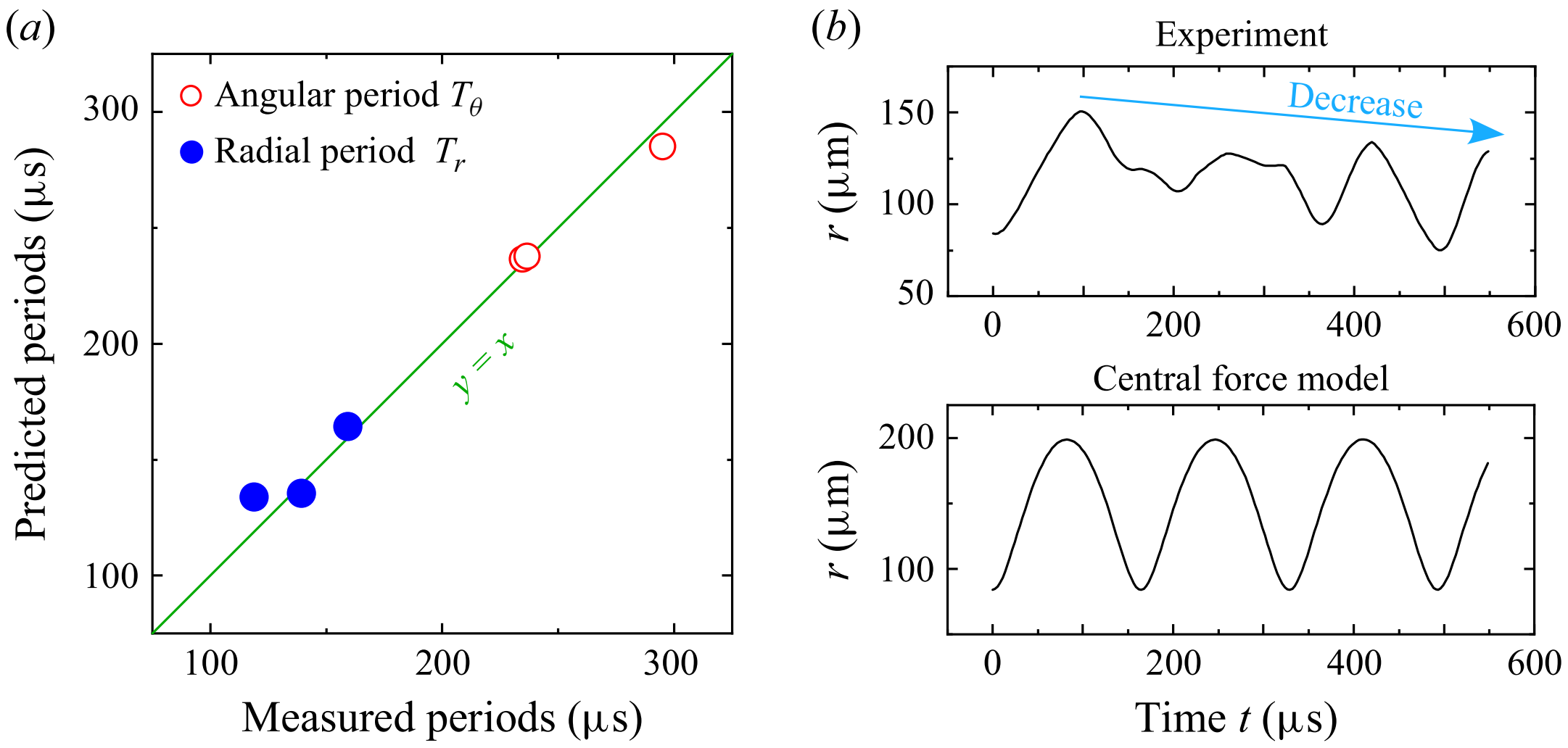

Figure 7. Comparison of the periods and radii between the central force model and the experimental results. (*a*) The angular period (red) and radial period (blue) from three different experiments are shown. The measured periods agree with the predicted periods. (*b*) The plots of $r(t)$ of the orbits in figures 6(*b*) and 6(*e*) are shown. The measured orbit is shrinking over time.

where $r(t)$ and $\theta(t)$ are the relative distance and angle at time $t$ in the polar coordinate system. The initial conditions, $v_r(t) \equiv r'(t)$ and $v_\theta(t) \equiv r(t)\theta'(t)$, are measured from experiments. The value of $v_\theta(0)$ is the measured Taylor–Culick speed, i.e. the constant film retraction speed before spinning. Open non-circular orbits are obtained as expected, according to Bertrand's theorem.

We experimentally measure the orbits using two high-speed cameras positioned at orthogonal angles, and then compare them with the theoretical orbits. The measured orbits are shown in figure 6(*b*–*d*). The dots with error bars represent measured data, and the line shows an interpolated orbit. The colours indicate time progression, ranging from red ($t = 0$) to purple ($t = 550$, 650 and 390 µs). We calculate the orbits, as shown in figure 6(*e*–*g*), by solving (3.15) numerically with real liquid properties and measured initial conditions (see figure captions). The experiments agree with numerical calculations semiquantitatively, revealing similar non-circular open orbits.

We compare the measured and predicted orbits by their periods, as shown in figure 7(*a*). Although bounded open orbits cannot be characterised by rotation period, they can instead be characterised by the angular period $T_\theta$ and the radial period $T_r$. We calculate the angular period by $T_\theta = \Delta t_\theta 2\pi/\Delta\theta$, where $\Delta\theta$ is the angular distance travelled during a time interval $\Delta t_\theta$. The radial period is calculated by $T_r = \Delta t_r/N_{cc}$ , where $N_{cc}$ is the number of crest-to-crest cycles observed in the plot of $r(t)$ over a time interval $\Delta t_r$ (Arya 1997, p. 258). The measured and predicted periods of three different experiments are shown in figure 7(*a*), which shows reasonably good agreement.

The main discrepancy is that the size of the measured orbit is smaller and shrinks over time, as highlighted by the downward trend of $r(t)$ in figure 7(*b*), which is inconsistent with the model. We believe the shrinking of the orbit is due to the air resistance and the variation in mass of the rims over time, which in turn results from the instabilities and the axial flow that will be discussed in the next section. Further studies on the internal fluid flow could improve the model's accuracy.

Although the analysis considers only the centre plane ($z = 0$, or equivalently, $\beta = 0$), the central force model should remain applicable to off-centre planes by recognising that the corresponding initial velocity is $v_\theta(0) = u_c \cos\beta$. Recall that the kinematic model in (3.2) and figure 2 has already assumed the ribbon is a ruled surface by connecting two rims, with a satisfactory result. However, it is challenging to compare with experimental results because the orbits in off-centre planes drift outward along the axial direction, which will be discussed in the next section.

This system presents an unusual scenario that the central force is distance-independent (potential is linear), in contrast to the typical celestial orbits (inverse-square law) and harmonic oscillators (Hooke's law). Other examples of systems with linear potential include the triangular quantum well in high-electron-mobility transistors (Harrison & Valavanis 2016, p. 116) and the quark confinement in mesons (Griffiths 2020, p. 173).

## 4. Corrugations and ligaments

### 4.1. *Phenomenology*

We observe that the rims become unstable during spinning at later times, forming corrugations along their lengths that grow into ligaments, which pinch off and eject secondary droplets, as shown in supplementary movie 7 and figure 8(*a*). The corrugations are marked by red dots. The ligaments are marked by cyan arrows. The instability is not uniquely associated with spinning, as similar corrugations are also observed without spinning. However, in the latter case, they do not grow into ligaments, as shown in figure 8(*b*).

We can distinguish this phenomenon from the previously discovered rim-splashing phenomenon by inspecting supplementary movie 7 and figure 8(*a*), even though they appear similar in still images (Néel *et al.* 2020; Tang *et al.* 2024). First, unlike in rim splashing, the two rims of the ribbon have not coalesced when the ligaments develop. Second, in rim splashing, the ligaments are straight and perpendicular to the film. For the spinning ribbon, in contrast, the ligaments are rotating with the ribbon. For example, the centre ligament in supplementary movie 7 has been rotated by $\sim 180^\circ$. Third, under spinning conditions, ligaments that break up into droplets are observed at a Weber number lower than that in rim splashing on a planar surface. For comparison, under the current spinning conditions, the local Weber number is $We_{loc} = \rho(2u_c)^2(2R_{rim})/\gamma \sim 58$, where $R_{rim}$ is the rim radius. In rim splashing, ligaments that break up into droplets are observed at $We_{loc} > 120$ (Néel *et al.* 2020; Tang *et al.* 2024). Although we have not yet explored the threshold Weber number, the data suggest that ligaments and droplets can be produced at a lower Weber number in this rim-spinning case compared with the rim-splashing case.

### 4.2. *Plateau–Rayleigh and Rayleigh–Taylor instabilities*

We plot the wavelength of the instability, $\lambda$, against the rim radius, $R_{rim}$, as shown in figure 8(*c*). The plot includes data for both spinning and non-spinning conditions for two different surface tensions. The average wavelength $\lambda$ is calculated by dividing the length of a segment by the number of corrugations $N$ it contains. The rim radii are measured either just before spinning begins or, for the non-spinning case, just after the two rims coalesce. For all conditions studied here, the data agree with the Plateau–Rayleigh instability that $\lambda_{PR} = 9.0 R_{rim}$ (Rayleigh 1878; Wang *et al.* 2018). Therefore, it is very likely that the Plateau–Rayleigh instability leads to the corrugations, while the spinning motion is also essential for the formation of ligaments that pinch off into droplets.

On the other hand, the corrugation may also be induced by Rayleigh–Taylor instability under the centripetal acceleration caused by spinning (Eisenklam 1964). The wavelength of Rayleigh–Taylor instability is $\lambda_{RT} = 2\sqrt{3}\pi l_c$, where $l_c = \sqrt{\gamma/(\rho a)}$ is the capillary length, $\gamma$ is the surface tension, $\rho$ is the liquid density, $a = R\omega^2$ is the centripetal acceleration, $R$ is the rotation radius, $\omega$ is the rotational speed.

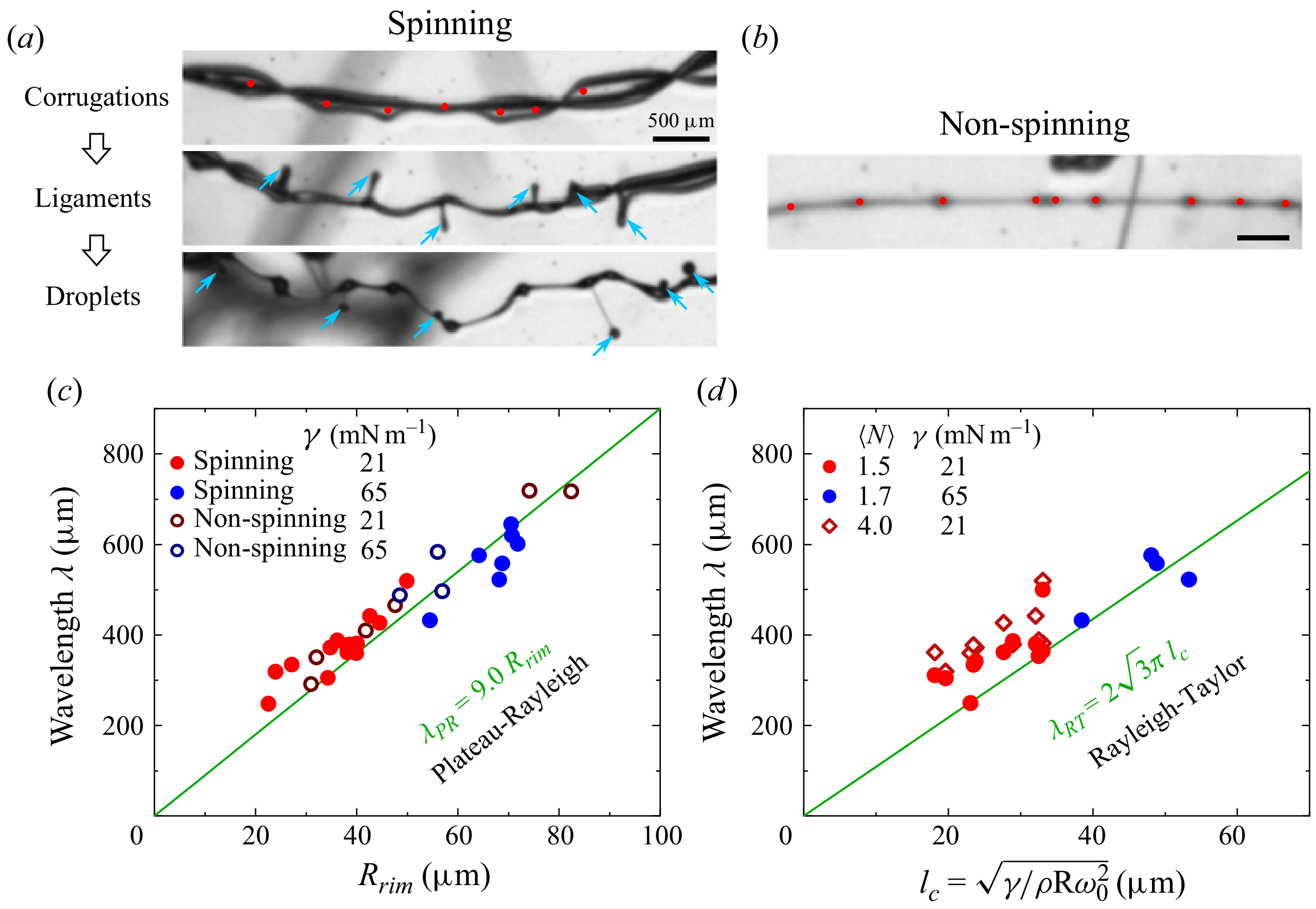

Figure 8. Corrugations and ligaments. (*a*) The corrugations on the rims are marked by the red dots. The ligaments are marked by the cyan arrows. Corrugations grow into ligaments, pinch off and then eject secondary droplets due to the spinning. (*b*) For the non-spinning case, the corrugations do not grow into ligaments. (*c*) The measured average wavelength $\lambda$ agrees with the Plateau–Rayleigh instability (solid line) for both spinning (solid dots) and non-spinning (open dots) conditions at different rim radius $R_{rim}$ and surface tensions $\gamma$. (*d*) The measured average wavelength $\lambda$ agrees with the Rayleigh–Taylor instability (solid line) at different capillary length $l_c$ and surface tensions $\gamma$. The number of segments used in the averaging is denoted by $N$. Because $l_c$ is calculated based on the centre rotational speed $\omega_0$, the average wavelength based on fewer segments is more reliable.

However, the rotational speed, and thus the centripetal acceleration, is not a constant but position dependent, along $z$. It is highest at the centre plane ($z = 0$) and decreases towards the two ends. Focusing on the region near the centre plane, we take $\omega \rightarrow \omega_0$ as the angular speed at $z = 0$ measured before the corrugations are observed. Correspondingly, the measured wavelengths are calculated using one or two segments nearest to the centre. The plot of the measured wavelengths versus the capillary length $l_c = \sqrt{\gamma/(\rho R\omega_0^2)}$ is shown in figure 8(*d*) and is compared with the theoretical expression (solid line). The experimental data is close to the expected values for the Rayleigh–Taylor instability.

In addition, we calculate another set of average wavelengths that includes more segments that are farther from the centre plane, as shown in figure 8(*d*) (open symbols). The average number of segments used is $\langle N \rangle = 4.0$. Nearly all of the obtained wavelengths increase. This agrees with the model that the rotational speed of the ribbon is highest at the centre plane and decreases towards the two ends, given that $\lambda_{RT} \sim 1/\omega$.

We have shown that the measured wavelengths agree with both the Plateau–Rayleigh and Rayleigh–Taylor instabilities. Comparing the wavelengths of Plateau–Rayleigh and Rayleigh–Taylor instabilities, one can show that $\lambda_{RT}/\lambda_{PR} = 1.2/\sqrt{Bo}$ where $Bo = \rho a R_{rim}^2/\gamma$ is the local rim Bond number, in which the rim radius $R_{rim}$ is taken as the characteristic length, and $a$ is the acceleration. It is commonly found that $Bo \sim O(1)$ in different breakup scenarios (Wang *et al.* 2018), so that $\lambda_{RT} \sim \lambda_{PR}$. In our experiments, calculated from the rotational speed at the centre plane, $Bo$ ranges from 1.5 to 4.6 so that the ratio $\lambda_{RT}/\lambda_{PR}$ ranges from 0.6 to 1.0. Therefore, the wavelengths of Plateau–Rayleigh and Rayleigh–Taylor instabilities are similar in magnitude under current experimental conditions.

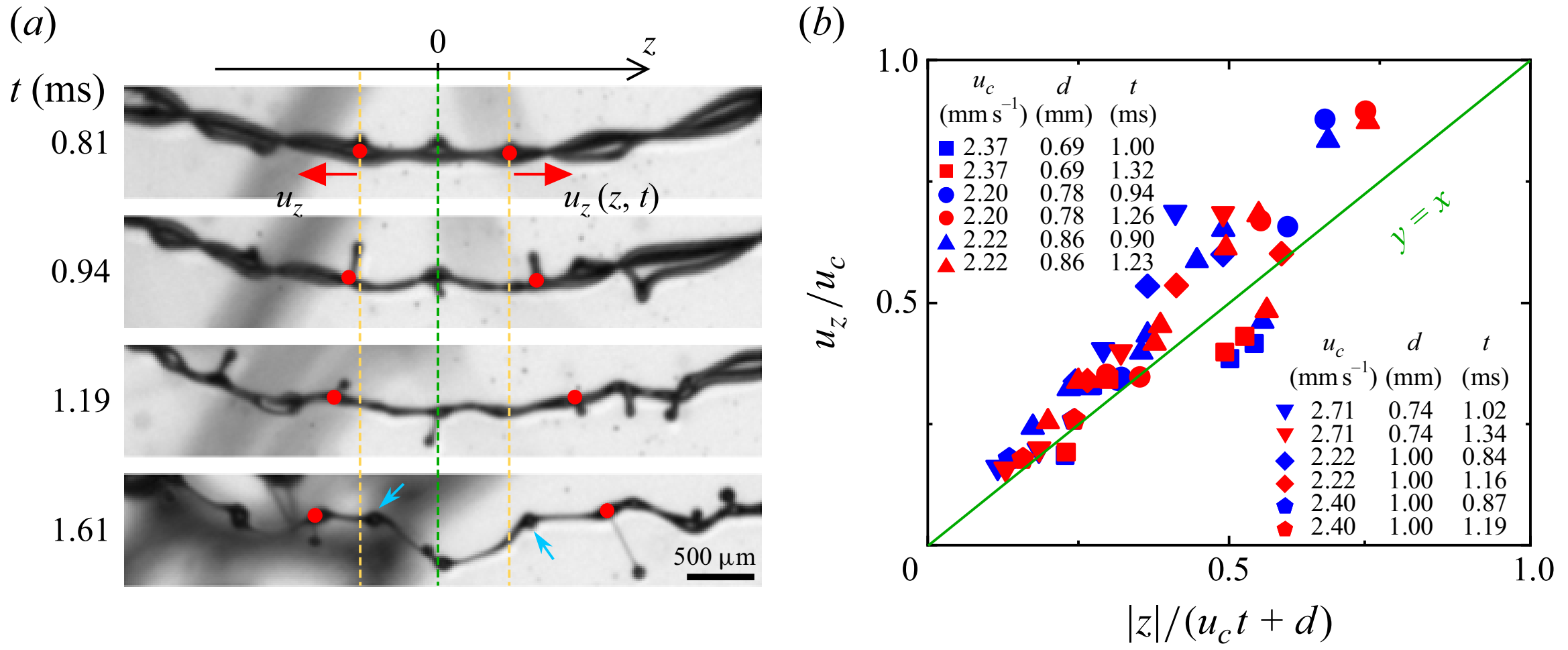

Figure 9. Outward axial flow. (*a*) Snapshots of the twisted ribbon at different times. The corrugations and ligaments (red dots) are moving outward with speed $u_z$ relative to the centre ($z = 0$) of the twisted ribbon, confirming the existence of the outward axial flow. The blue arrows highlight the newly emerged corrugations at later times. (*b*) The speed of the outward axial flow $u_z$ measured at different conditions as shown in the legend. See also supplementary movie 7.

### 4.3. *Axial flow*

The corrugations and ligaments are moving outward with speed $u_z$ relative to the centre ($z = 0$) of the twisted ribbon, as indicated in figure 9(*a*), confirming the existence of the outward axial flow. It originates from the rims' non-zero velocity component parallel to the rotation axis, as the rims move radially from the points of rupture before reaching the rotation axis. By (3.2), the axial flow speed $u_z$ is given by

$$\frac{u_z(z, t)}{u_c} = \sin\beta = \frac{|z|}{u_c t + d}. \tag{4.1}$$

This relation agrees with experimental data as shown in figure 9(*b*). Experimentally, $u_z$ is measured by tracking the positions of corrugations over time, and $d$ is measured as the distance between the ribbon and the smaller hole without considering the asymmetry.

It is interesting that, by volume conservation, the outward flow would reduce the thickness of the sheet and the rim, adding complexity to the understanding of the rims' orbit. Unfortunately, the resolution in our experiments is insufficient to quantify this change in thickness directly.

The axial flow also causes the emergence of new corrugations over time, as shown in the last row of figure 9(*a*) and supplementary movie 7. Due to the axial flow, the spacing between neighbouring corrugations or ligaments increases over time. When the spacing becomes sufficiently large relative to the wavelength of the instability, new corrugations begin to emerge. Such self-sustained population of corrugations has been studied previously in the rim of an expanding liquid sheet (Gordillo, Lhuissier & Villermaux 2014).

## 5. Conclusions

To conclude, we have explained the formation and evolution of spinning twisted ribbons. The ribbon forms when the rim deviates from the original curved surface due to insufficient centripetal force. This phenomenon is unique to curved films and does not occur in planar films. The geometry and motion of the ribbons are described by the kinematic model, the asymmetric kinematic model and the central force model. The calculated surfaces resemble the ribbons observed in the experiment. The positions of the conjunction points are accurately predicted. The mean curvature of the surface is small. The orbit of the rim is open and non-circular. Due to the spinning, corrugations along the rims grow into ligaments that eventually pinch off, ejecting secondary droplets. These corrugations very likely arise from the Plateau–Rayleigh and/or Rayleigh–Taylor instabilities, as the measured wavelengths agree with both models. This rim-spinning phenomenon is distinctly different from the rim splashing observed in previous studies, including its facilitation of droplet formation at lower Weber numbers. The ribbon contains an intrinsic outward axial flow, which leads to the emergence of new corrugations and variations in rim thickness. While this study focuses on experiments of multiple-hole rupture in corona splash, the underlying principles are likely applicable to other systems where twisted ribbons emerge.

**Supplementary movies.** Supplementary movies are available at https://doi.org/10.1017/jfm.2025.10299.

**Acknowledgements.** We acknowledge V. Mugundhan, A. Aguirre-Pablo, K. Dharmarajan, M. Lin, Z. Yang, A. Alhareth, F. Kamoliddinov and M. Kattoah for their help in experiments and fruitful discussions. We thank the referees for their thoughtful comments and valuable suggestions.

**Funding.** This work is financially supported by King Abdullah University of Science and Technology (KAUST) under grant numbers URF/1/2621-01-01 and BAS/1/1352-01-01.

**Declaration of interests.** The authors report no conflict of interest.

**Data availability statement.** All data is included in the manuscript and supplementary movies.

**Author contributions.** S.T.T. conceived the project; T.A., J.H.Y.L. and Y.L. built the experimental set-up; J.H.Y.L., Y.L. and S.T.T. performed the experiments; J.H.Y.L., Y.L. and S.T.T. analysed the data; J.H.Y.L. and Y.L. constructed the models; J.H.Y.L. and M.F.A. performed the numerical simulations; J.H.Y.L, Y.L. and S.T.T. wrote the manuscript.

## Appendix A. Crown sheet composition

To confirm that the crown of the corona splash (i.e. the curved film) originates from the liquid in the drop, under our experimental conditions, rather than from the liquid film coating the glass slides, we conduct dyed-drop experiments, as shown in figure 10. The drop is a glycerol–water mixture dyed yellow with fluorescein (0.01wt%). From figure 10, we can see that the crown shows the same yellow colour as the drop.

## Appendix B. Correction of parallax error

On a curved surface, the plane of ruptures may not be parallel to the image plane of the camera, inducing errors in length measurements. We correct these parallax errors by the following method.

First, we use the side-view image to measure the local rotation angles (Euler angles) $\theta$ and $\phi$, as shown in figure 11. The angle $\phi$ is read from the image directly, while the angle $\theta$ is obtained, assuming axisymmetry, from the equation $\sin\theta = s/R_h$, where $R_h$ is the horizontal radius of the crown, $s$ is the distance of the plane from the centre. A top-view drawing is provided in figure 11 for clarity.

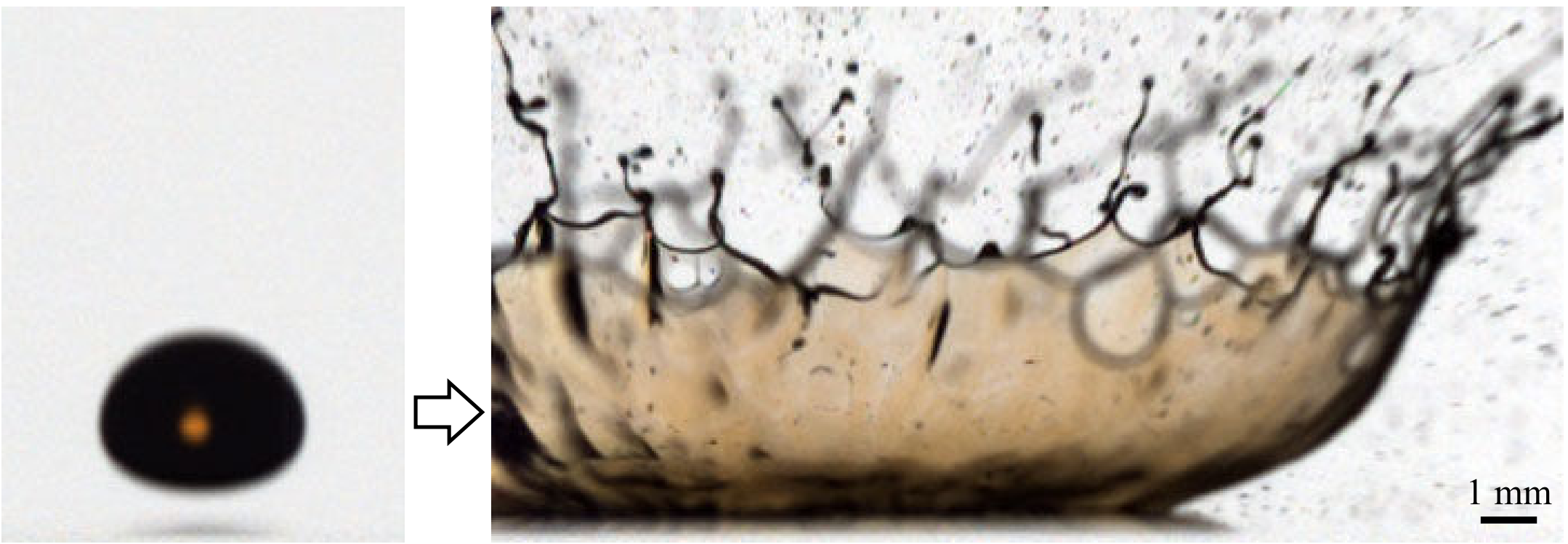

Figure 10. Dyed-drop experiment indicates that the crown sheet originates from the liquid in the drop.

Next, we derive the formula to correct for the parallax errors in length measurements. The rotation matrix is

$$\boldsymbol{R} = \begin{pmatrix} 1 & 0 & 0 \\ 0 & \cos\phi & -\sin\phi \\ 0 & \sin\phi & \cos\phi \end{pmatrix} \begin{pmatrix} \cos\theta & 0 & \sin\theta \\ 0 & 1 & 0 \\ -\sin\theta & 0 & \cos\theta \end{pmatrix} = \begin{pmatrix} \cos\theta & 0 & \sin\theta \\ \sin\theta\sin\phi & \cos\phi & -\cos\theta\sin\phi \\ -\cos\phi\sin\theta & \sin\phi & \cos\theta\cos\phi \end{pmatrix}, \tag{B1}$$

while the projection matrix is

$$\boldsymbol{P} = \begin{pmatrix} 1 & 0 & 0 \\ 0 & 1 & 0 \\ 0 & 0 & 0 \end{pmatrix}. \tag{B2}$$

Let $\boldsymbol{d} \equiv \begin{pmatrix} d_x \\ d_y \\ 0 \end{pmatrix}$ be an arbitrary unit line on a plane, then $\boldsymbol{d}' \equiv d' \begin{pmatrix} \cos\psi \\ \sin\psi \\ 0 \end{pmatrix}$ is the rotated and projected line from $\boldsymbol{d}' = \boldsymbol{PRd}$, where the angle $\psi$ is measured from the x-axis. We then get

$$\begin{aligned} d_x &= d' \cos\psi \sec\theta, \\ d_y &= d' \left(\sin\psi \sec\phi - \cos\psi \tan\theta \tan\phi\right). \end{aligned} \tag{B3}$$

Therefore, the ratio of the length before and after this transformation is

$$\Psi(\psi) = \frac{|\boldsymbol{d}|}{|\boldsymbol{d}'|} = \frac{d_x^2 + d_y^2}{d'} = \sqrt{\left(\frac{\cos\psi}{\cos\theta}\right)^2 + \left(\frac{\sin\psi}{\cos\phi} - \cos\psi\tan\theta\tan\phi\right)^2}, \tag{B4}$$

where $\Psi$ is the ratio between the real length, $d$, and the apparent length, $d'$. In other words, we can deduce the real length from the apparent length by $d = \Psi d'$.

To verify (B4), we test it on the diameter of a hole. We make use of the facts that (i) physically, the capillary-driven hole is circular, and (ii) the apparent length of the longest axis of the transformed hole, $D'_{max}$, is the real diameter of the hole, $D$. Experimentally, we measure $D'_{max} = D$ and the width of the hole along an arbitrary axis, $D'$, as shown in figure 12. We calculate the ratio $D/D'$ and compare it with the ratio $\Psi$ deduced by (B4). In the example in figure 12, for $\psi = 59.5^\circ$, $\theta = -33.6^\circ$, $\phi = 27.5^\circ$, we get $D/D' = 1.256$ and $\Psi = 1.259$. The measured ratio and the calculated ratio are very close, verifying that (B4) is correct.

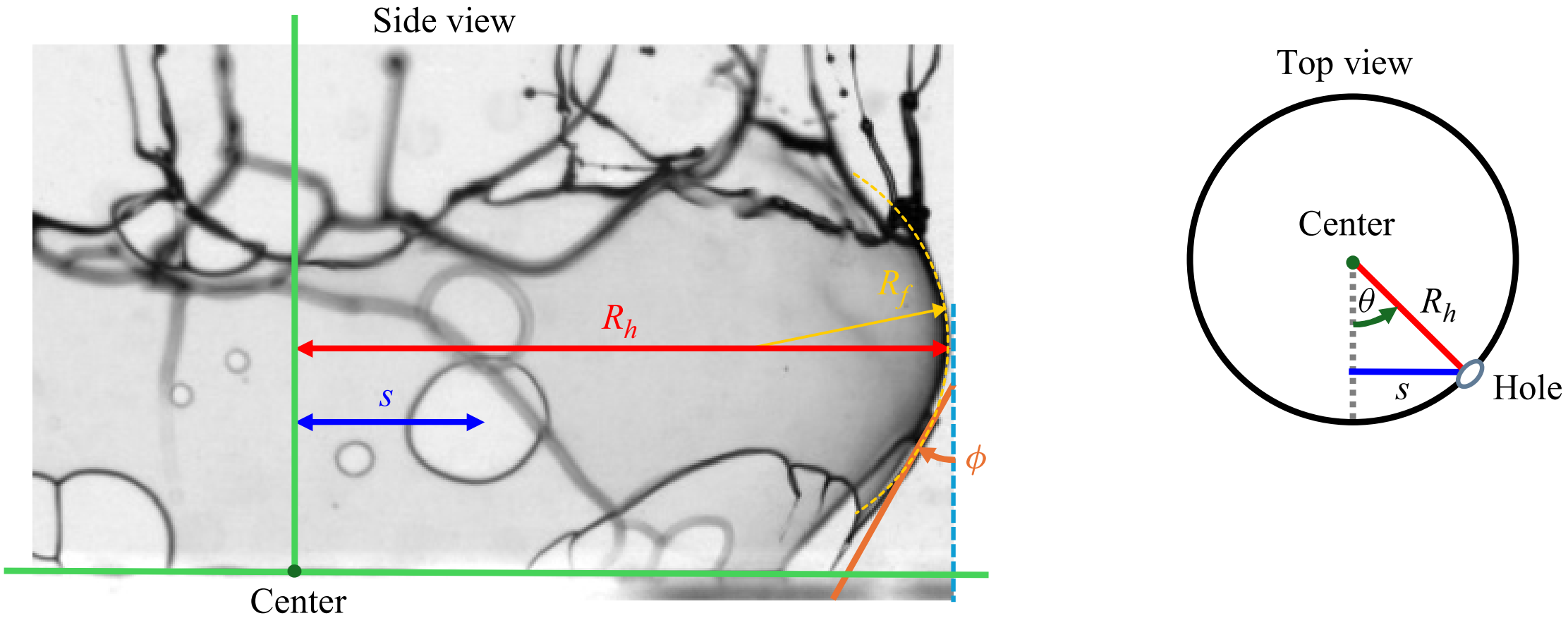

Figure 11. Side and top views for parallax corrections.

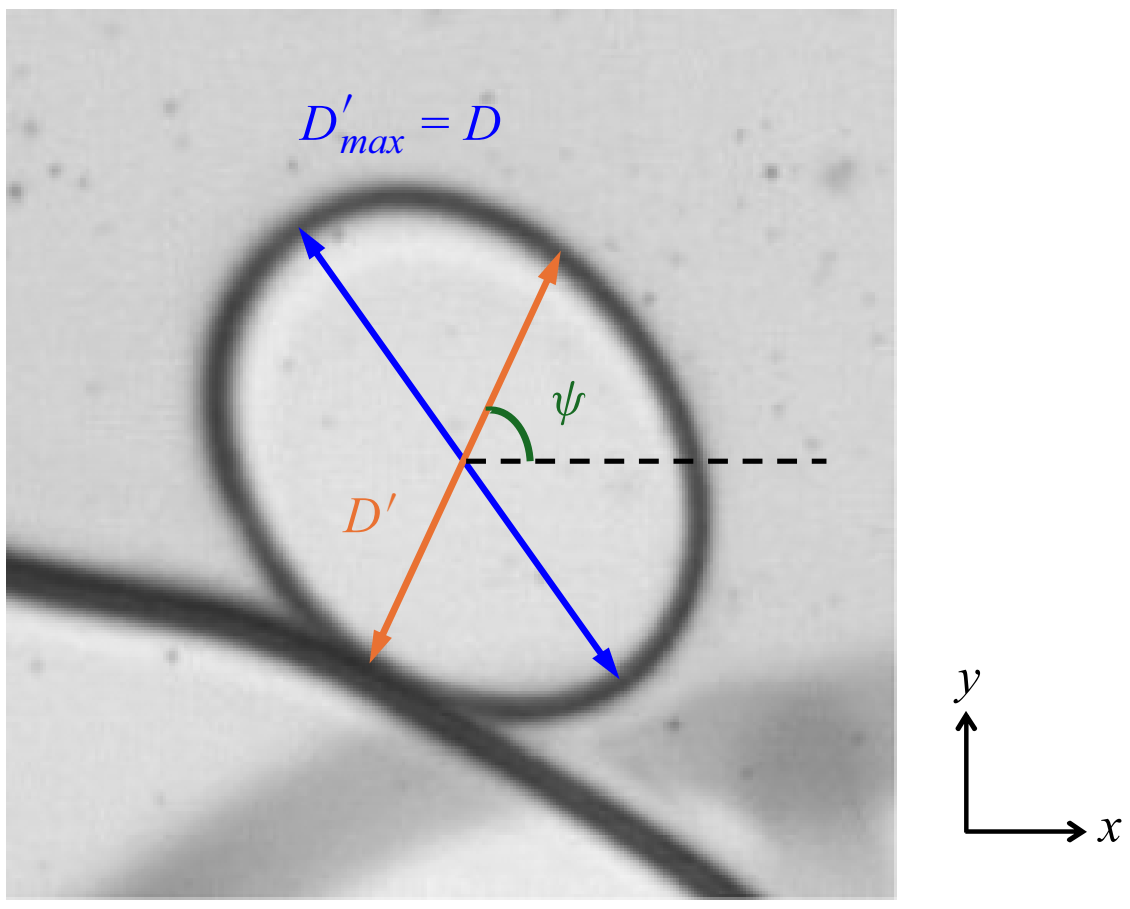

Figure 12. Image used to verify the parallax correction and (B4).